\documentstyle[psfig,emulateapj]{article}

\def\etal{et al.\,}

\def\refnew#1{(\ref{#1})}
\def\be{\begin{equation}}
\def\ee{\end{equation}}

\def\erg{\, \rm erg}

\def\s{\, \rm s}
\def\km{\, \rm km}
\def\cm{\, \rm cm}

\def\m{\, \rm m}

\begin{document} 
	
\title{\mbox{Planet Migration and Binary Companions: the case of HD
$80606$b}}

\lefthead{Migration \& Binarity}
\righthead{Wu \& Murray}

\author{Y. Wu\altaffilmark{1} \&  N. Murray\altaffilmark{1,2}}

\altaffiltext{1}{Canadian Institute of Theoretical Astrophysics,
	University of Toronto, 60 St. George Street, Toronto, Ontario
	M5S 3H8, Canada}
\altaffiltext{2}{Canada Research Chair in Astrophysics}

\begin{abstract}

The exo-solar planet HD $80606b$ has a highly eccentric ($e=0.93$) and
tight ($a=0.47$ AU) orbit. We study how it might arrive at such an
orbit and how it has avoided being tidally circularized until now. The
presence of a stellar companion to the host star suggests the
possibility that the Kozai mechanism and tidal dissipation combined to
draw the planet inward well after it formed: Kozai oscillations
produce periods of extreme eccentricity in the planet orbit, and the
tidal dissipation that occurs during these periods of small pericentre
distances leads to gradual orbital decay. We call this migration
mechanism the 'Kozai migration'.  It requires that the initial planet
orbit is highly inclined relative to the binary orbit.  For a
companion at $1000$ AU and an initial planet orbit at $5$ AU, the
minimum relative inclination required is $\sim 85 \deg$.  We discuss
the efficiency of tidal dissipation inferred from the observations of
exo-planets.  Moreover, we investigate possible explanations for the
velocity residual (after the motion induced by the planet is removed)
observed on the host star: a second planet in the system is excluded
over a large extent of semi-major axis space if Kozai migration is to
work, and the tide raised on the star by HD $80606$b is likely too
small in amplitude. Lastly, we discuss the relevance of Kozai
migration for other planetary systems.

\end{abstract}


\setcounter{equation}{0}

\section{INTRODUCTION}
\label{sec:nl-intro}


Ongoing radial-velocity surveys have uncovered $\sim 100$ exo-solar
giant planets.\footnote{See http://cfa-www.harvard.edu/planets/ for a
constantly updated catalog.} Their orbital characteristics are
puzzling: many have eccentricities much higher than the planetary
orbits in our solar system, and a large fraction reside very close to
the host star. These orbits indicate either a formation scenario
different from that which operated in our solar system, or a migration
mechanism to bring the planets in from large distances.
 
The object HD $80606$b (Naef, Latham \& Mayor \etal, \cite{naef},
hereafter NLM, period $111.8$ days, $ m_p \sin i = 3.4
M_J$)\footnote{The host star HD $80606$ is a G5 star with [Fe/H] $=
0.43$, $m_B= 9.65$ and $m_V= 8.93$.}
the focus of this article, has the highest eccentricity ($e = 0.927\pm
0.012$) and the smallest pericentre distance ($\sim 7 R_\odot$) among
all known exo-planet candidates. Is it possible to form such a system
in-situ?
Among the spectroscopic binaries contained in the Batten catalog
(Batten, Fletcher \& MacCarthy \cite{batten}), the only G-type binary
that has a comparably small periastron distance ($\sim 6 R_\odot$) is
HD 27935 (Griffin, Griffin \& Gunn \etal, \cite{griffin}) with $e=0.85$.
Duquennoy, Mayor \& Andersen \etal (\cite{Gl586}) reported a more
eccentric system, the K-dwarf binary Gl 586A, with $e= 0.975$ and a
periastron distance of $10 R_\odot$. However, doubts exist as to
whether the latter binary can be primordial. But assuming both systems
are primordial, their presence empirically suggests that HD $80606$b
might have formed as a binary companion to HD $80606$.
This hypothesis is currently untestable, 
the more so since the projected mass-ratio in this system is rather
extreme.
Instead we ask the following question: can HD $80606$b be formed in
the cool outer part of a protoplanetary disk and then evolve into the
current orbit via migration?

Planet-planet scattering is one such possibility, assuming HD $80606$
had or has multiple planets.
Ford, Havlikova \& Rasio (\cite{ford}) integrated the evolution of
$\sim 10^3$ two-planet systems that are initially dynamically
unstable. Not a single close encounter produces a planet with $e >
0.90$.
This rarity leads us to conclude that either the orbits of most
observed exo-planets are results of planet-planet scattering, or that
the HD $80606$ system is not the result of such scattering.

The leading hypothesis for planet migration involves gravitational
interactions between the planets and the gas disks out of which they
form (Goldreich \& Tremaine \cite{gt}; Lin, Bodenheimer \& Richardson
\cite{lbr}). Under certain circumstances planet-disk interactions may
also excite the orbital eccentricity of the planet (see, e.g.,
Goldreich \& Sari \cite{goldsari}).
If so, this may explain the orbital separation and eccentricity of the
bulk of the systems discovered so far. These systems exhibit an
eccentricity distribution that is roughly flat below $e \sim 0.6$ and
distinctly drops off around $0.6$. It is believed that Jupiter-mass
planets can open up gaps in their natal gas disks with a fractional
width $\Delta a/a$ reaching up to a similar value. This coincidence
suggests that passage of an eccentric planet through the disk on
either side of the gap tends to damp the planet's orbital eccentricity
and to limit the maximum eccentricity planet-disk interaction produces
to $e\sim 0.6$. However, HD $80606b$ stands out with $e=0.93$.
%
%

The origin of the orbit of HD $80606$b becomes more perplexing if we
consider tidal effects.  Assuming that the planet has a tidal
dissipation efficiency similar to other known exo-planets (quality
factor $Q_p= 3\times 10^5$, Wu \cite{wu}) or to Jupiter (Goldreich \&
Soter \cite{goldreich}, Peale \& Greenberg \cite{peale}) and setting
its true mass equal to the minimum mass, its radius to a Jupiter
radius, we find that a mere $650$ Myrs ago, HD $80606$b had an
implausibly high eccentricity of $e = 0.99$.\footnote{
This problem is further exasperated if the planet has acquired the
same periastron distance while it was still hot and large.} The age
of the system is likely to be this old or older; 
%
the stellar projected rotational velocity is low, $v\sin i \approx 0.9
\km/\s$, typical of old G-dwarves; it is also chromospherically quiet.

Interestingly, there is a neighbor to this system, HD $80607$, a
main-sequence companion $\sim 1000$ AU away. This prompts us to
develop a theory in which the companion is responsible for the high
eccentricity and small orbit of the planet. There are two ingredients
in our theory. First, we assume that the planet was born on an orbit
of a few AU, and had an orbital plane inclined relative to the stellar
binary plane. The remote stellar companion would induce an
eccentricity oscillation in the planet's orbit via the Kozai mechanism
(Kozai\cite{kozai}; see also discussions by Holman, Touma \& Tremaine
\cite{holman}, Innanen, Zheng, \& Mikkola \etal, \cite{innanen}, and
Mazeh, Krymolowski \& Rosenfeld \cite{mazeh}).
%
The second component is tidal circularization. The tides operate most
effectively during episodes of high eccentricity in the Kozai
cycle. We rely on the dissipative tidal process to irreversibly draw
the planet inward. Combining the two processes, we can explain the
current high eccentricity as well as the tight orbit.  We refer to
this planet migration scenario as `Kozai migration'.

Eggleton, Kiseleva \& Hut (\cite{EKH}), Kiseleva, Eggleton \& Mikola
(\cite{kem}), and Eggleton \& Kiseleva-Eggleton (\cite{EKE}, EKE
hereafter), have developed this scenario to explain some hierarchical
triple star systems.  Blaes, Lee \& Socrates (\cite{omer}) applied an
analogous scheme (gravitational radiation being the dissipation
mechanism) to hierarchical triple black holes in galactic centers. In
this article, we adopt the formalism of EKE in an attempt to produce a
plausible life-history for HD $80606$b (\S
\ref{sec:HD80606}). The story depends on the value of the tidal
quality factor $Q$ (\S \ref{subsec:Qvalues}).  The observational
consequences of such a life-history follow in \S
\ref{subsec:secondplanet}. In \S \ref{subsec:tidalheight} we discuss
the tidal velocities induced at the stellar photosphere by the highly
eccentric planet, and the possibility of observing such tides. NLM
note the presence of substantial residuals to the Keplerian fit in the
HD $80606$ system. The tidal velocities appear to be large, but not
large enough to explain the observed velocity residuals. Lastly (\S
\ref{subsec:othersystem}), we assess the importance of Kozai migration
for other planetary systems where binary companions are also known to
exist.

\section{Kozai Migration for HD $80606$b}
\label{sec:HD80606}

The host star of HD $80606$b is known to be a member of a common
proper-motion binary with a companion (HD $80607$) that is similar to
HD $80606$ in both spectral type and metallicity . The two stars are
separated by $\sim 20''$ on the sky. Unfortunately, it is hard to
translate this into a linear dimension. The Hipparcos distances for
both stars are highly uncertain\footnote{Hipparcos parallaxes are $17
\pm 6$ mas for HD $80606$ and $10 \pm 9$ for HD $80707$. The large
error bars are presumably due to mutual contamination (see also NLM).} 
This complicates an age determination with isochrone fitting. But if
we enforce the constraint that the stars have reached the
main-sequence, isochrone fitting yields a lower limit of $\sim 65$ pc
on the distance, implying a projected binary separation of $1300$ AU
at the current epoch. On the other extreme, requiring that the stars
have not yet ascended the red giant branch yields a maximum distance
of $\sim 125$ pc, and a projected separation of $2500$ AU. As noted
above, the low $v\sin i$ and the chromospheric quietness of HD $80606$
are typical of an old G-dwarf.  In this article, we adopt a value of
$1000$ AU for the binary semi-major axis ($a_c$). The eccentricity is
taken to be $e_c = 0.5$.\footnote{We denote quantities related to the
planet and the companion star with subscripts $p$ and $c$,
respectively.} We discuss the effects on the Kozai migration when
$a_c$ is increased.

Even at such a large distance, the companion star could significantly
perturb the planet orbit as long as the two orbital planes are
initially inclined to each other more than $39.2 \deg$.\footnote{Below
this value, the periapse argument of the planet circulates through all
angles very fast and the torque from the companion is effectively
averaged to zero. There is little variation in the planet's
inclination and eccentricity. Also see, e.g., Holman \etal
(\cite{holman}).}
Secular effects of the Kozai type could then occur and produce large
cyclic variations in the planet's eccentricity ($e_p$) and relative
inclination, as a result of angular momentum exchange with the
companion orbit. Effects on the companion orbit produced by the planet
can be neglected, given the extreme ratio in the two orbits' angular
momenta ($a_c \gg a_p$ and $m_c \gg m_p$).  Under this approximation,
the z-component of the planet's angular momentum, $J_z = \sqrt{G
M_\star a_p} M_p \sqrt{1-e_p^2} \cos I$, is conserved.\footnote{See
Ford, Kozinsky \& Rasio \cite{ford0} for a discussion of higher order
effects).} Here the z-axis is normal to the companion plane.  As $a_p$
is not modified by secular perturbations, the Kozai integral $H_K =
\sqrt{1-e_p^2} \cos I$ is conserved during the oscillations. Minima in
$e_p$ concur with maxima in $I$, and {\it vice versa}. Each Kozai
cycle lasts a time $\sim P_c^2/P_p$, or a few Myrs.

A much slower process, tidal circularization, gradually removes energy
from the orbit and draws the planet inward.  An initially more highly
inclined orbit (larger $I$) could reach higher eccentricity. And since
the dissipation process depends sensitively on the nearest approach
distance between HD $80606$b and its host star, such an orbit would
suffer stronger orbital decay and have shorter migration timescale.
The key issue in Kozai migration is therefore the initial value of
$I$. In \S \ref{subsec:heuristic}, we first
estimate the minimum initial inclination required for migrating HD
$80606$b from an initial orbit of $a_p = 5$ AU and an initial
eccentricity of $e_p = 0.1$, we then confirm this estimate numerically
in \S \ref{subsec:numerical}.

\begin{table*}
\begin{tabular}{ccr}
\hline
Symbol & Definition & Fiducial Values \\
\hline

$M_\star, M_c, M_p$ & stellar (host \& companion) and planetary masses &
$1.1 M_\odot$, $1.1 M_\odot$, $7.80 M_J$\\

$R_\star, R_p$ & stellar and planetary radii & $R_\odot$, $R_J$ \\

$k_{2\star}, k_{2p}$ & tidal Love number & $0.028$, $0.51$ \\

$r_{g\star}, r_{gp}$ & gyro-radius, moment of inertia $ = r_g M R^2$
 & $0.08$, $0.25$ \\

$Q_\star, Q_p$ & tidal dissipation quality factors & $10^6$,
$3\times 10^5$ \\

$\Omega_\star, \Omega_p$ & spin frequency & initially $20$ days \&
$10$ hours\\


$a_p$, $e_p$, $n_p$ \& $P_p$ & planet semi-major axis, eccentricity,
mean motion \& period & $n_p = \sqrt{G(M_\star + M_p)/a_p^3}$ \\

$a_c$, $e_c$, $n_c$ \& $P_c$ & parameters for the companion orbit & $a_c =
1000$ AU, $e_c = 0.5$ \\

\hline
\end{tabular}
\label{table:symbol}
\end{table*}

\subsection{Minimum Initial Inclination: heuristic argument}
\label{subsec:heuristic}

To estimate the minimum initial inclination, we note that while the
Kozai oscillation roughly conserves $\sqrt(1-e_p^2) \cos I$ and $a_p$
during individual cycles, tidal circularization preserves orbital
angular momentum ($J= \sqrt{G M_\star } M_p \sqrt{a_p (1-e_p^2)} $)
during episodes of maximum eccentricity. The planetary spin is quickly
(pseudo-)synchronized 
with the orbital motion,\footnote{Pesudo-synchronization applies to an
eccentric system where the planet spin is tidally synchronized to a
rate that is in-between the orbital frequency and the angular
frequency at periapse. See, e.g., Hut (\cite{hut}).}
while the star's synchronization time greatly exceeds the orbital
circularization time. Hence the planet's orbital angular momentum
cannot be absorbed into spin. Since $e_p \sim 1$ near maximum,
conservation of $J$ translates into a constant periastron distance
between maximum $e_p$ during the Kozai cycles.

In an orbit with a semi-major axis of $5$ AU, HD $80606$b would have
to reach an eccentricity as high as $0.993$ (at some point during the
Kozai cycle) in order to produce the currently observed
periastron. Adopting a minimum of $I \approx 40 \deg$ during the Kozai
cycles (see, e.g., Holman \etal, \cite{holman}), the Kozai integral (
$\sqrt(1-e_p^2) \cos I$) remaining constant yields $I \geq 84.8 \deg$
for an initial $e_p=0.1$.  In other words, the two orbits had to be
virtually perpendicular to each other when the planet formed.
This conclusion is independent of the companion orbital separation.

The planet in HD $80606$b is not currently undergoing Kozai
oscillations; the Kozai mechanism operated only when the planet had a
semi-major axis $a\geq 4$ AU.  To see why, note that the torque
exerted by the companion star on the planet has its value and
direction dependent on the pericentre argument of the planet
($\omega_p$). If, however, this argument also precesses under other
forces, the averaged Kozai torque is reduced.  General relativistic
effects, tidal effects, and rotational quadrapolar bulges on the
planet\footnote{Analogous but less important bulges on the star are
ignored.} are mainly responsible for these extra precession. Their
rates are summarized here (Sterne \cite{sterne}, Einstein
\cite{einstein}), with definitions for the symbols listed in Table 1.,
\begin{mathletters}
\begin{eqnarray}
\hskip-0.5cm {d\omega_p\over dt}\Bigr|_{\rm GR} & = & 3
n_p\, {{G M_\star}\over{a_p c^2 (1-e_p^2)}},\label{eq:preGR} \\
\hskip-0.7cm {d\omega_p\over dt}\Bigr|_{\rm rot} \,\,& = & {1\over2}
n_p {k_{2}\over{(1-e_p^2)^2}}\!\left({{\Omega_p}\over{n_p}}\right)^2\!\!
{{M_\star}\over{M_p}}\!\!  \left({{R_p}\over{a_p}}\right)^5,
\label{eq:preJ2}\\
\hskip-0.7cm {d\omega_p\over dt}\Bigr|_{\rm tide} & = & {15\over2} n_p
k_{2}\! {{1+\frac{3}{2}e_p^2 + \frac{1}{8}e_p^4}\over{(1-e_p^2)^5}}
{{M_\star}\over{M_p}}\!\!  \left({{R_p}\over{a_p}}\right)^5.
\label{eq:pretide}
\end{eqnarray}
\end{mathletters}
Taking a Kozai precession rate of $\sim 0.5 n_c^2/n_p/(1-e_c^2)^{3/2}$
(Holman \etal, \cite{holman}), and assuming that the planet is
pseudo-synchronously spinning with the orbit ($e_p = 0.1$), the
relative precession rates are,
\begin{eqnarray}
{{\dot{\omega}_p|_{\rm GR}}\over{\dot{\omega}_p|_{\rm Kozai}}}&
\approx & 5900
\left({{a_1}\over{0.47 {\rm AU}}}\right)^{-4}\!\!
\left({{a_c}\over{1000 {\rm AU}}}\right)^{3}, \nonumber\\
{{\dot{\omega}_p|_{\rm rot}}\over{\dot{\omega}_p|_{\rm Kozai}}}&
\approx & 750
\left({{a_1}\over{0.47 {\rm AU}}}\right)^{-8}\!\!
\left({{a_c}\over{1000 {\rm AU}}}\right)^{3}\!\!
\left({{R_p}\over{R_J}}\right)^{5},  \nonumber \\
{{\dot{\omega}_p|_{\rm tide}}\over{\dot{\omega}_p|_{\rm Kozai}}}&
\approx & 270
\left({{a_1}\over{0.47 {\rm AU}}}\right)^{-8}\!\!
\left({{a_c}\over{1000 {\rm AU}}}\right)^{3}\!\!
\left({{R_p}\over{R_J}}\right)^{5}\!\!\!\!.
\label{eq:precession}
\end{eqnarray}
If any of the above ratios exceeds unity, the Kozai oscillation is
destroyed. Setting the first ratio to unity, we obtain a suppression
radius $\approx 4$ AU. Currently, the Kozai oscillations are strongly
suppressed by these extra precessions. This conclusion does not affect
the minimum inclination required for Kozai migration. However, the
suppression radius does depend on the companion separation: e.g., it
is reduced to $\approx 1.5 $AU if the companion semi-major axis is as
small as $250$ AU.


\subsection{Numerical Integration}
\label{subsec:numerical}

We adopt the set of equations (eq. [11]-[17]) in EKE (also see
Eggleton, Kiseleva \& Hut \cite{EKH}) to describe the secular
evolution of the planet orbit, as well as the spin of the host star
and the planet. Any change in the companion's orbit is
ignored. Physical effects described by these equations include:
secular interactions, GR, tidal and rotational precession, and tidal
dissipation. These equations are explicitly listed in Appendix
\ref{sec:EKEeqn}. 

We integrate equations \refnew{eq:ep}-\refnew{eq:omegapi} starting
from the following initial conditions: planet orbit $a_p =5.0$ AU,
$e_p = 0.1$, relative inclination $I = 85.6\deg$ (slightly higher
than the $84.8\deg$ estimate in \S \ref{subsec:heuristic} because the
actual minimum inclination is $\sim 48\deg$, not $\sim 40\deg$ as we
assumed), and periapse angle $\omega_p = 45\deg$.

\begin{figure*}[t]
\centerline{\psfig{figure=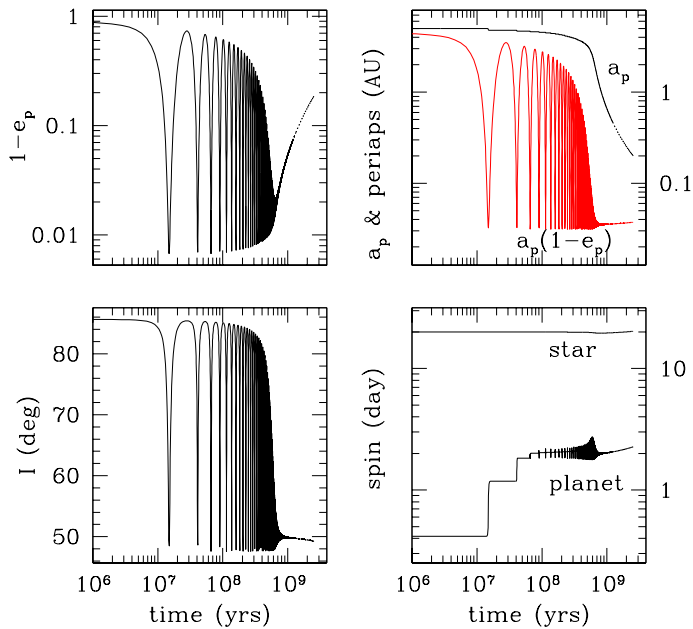,width=0.50\hsize}}
\caption{Evolution of the planetary orbit as a function of logarithmic time
(in years). The four panels represent (from left to right, up to
down), evolution of the planet eccentricity, planet semi-major axis
and periastron distance (in AU), inclination of the planet plane
relative to the stellar binary plane, and spin periods for both the
planet and the star (in days).  See the text for the choice of
parameters. During the Kozai cycles, maxima in eccentricity concur
with minima in inclination. At the current orbit ($e_p = 0.927$ and
$a_p = 0.47$ AU), the planet no longer undergoes Kozai oscillations as
a result of GR and tidal precession. The dashed part of each curve
depicts future evolution of the planet orbit under tidal
circularization.}
\label{fig:eggleton_short} 
\end{figure*}

Values of the system parameters are listed in Table 1.
The resulting orbital and spin information are presented in
Fig. \ref{fig:eggleton_short} as functions of time. Initially, the
planet undergoes Kozai oscillation with each cycle lasting $\sim 20$
Myrs. Whenever $e_p$ evolves close to maximum, a small amount of
energy (but not angular momentum) is removed from the orbit. Both
$a_p$ and $e_p$ are decreased,
while $a_p(1-e_p)$ remains roughly constant.  This corresponds to a
small kick to reduce $e_p$ whenever $\omega_p
\approx \pi/2$ (see Fig. 3 of Holman \etal \cite{holman}).
The Kozai integral slowly increases. This explains why the amplitude
of $e_p$ oscillation shrink as the planet migrates inward (top-left
panel of Fig. \ref{fig:eggleton_short}). After about $0.7$ Gyrs, the
planet has reached an orbit with $a_p \sim 2.5$ AU, and the
Kozai oscillation is completely destroyed by GR precession. At this
epoch, we find $e_p \sim 0.99$ and $I \sim 50 \deg$.  From this point
on, tidal circularization dominates the evolution, and the influence
of the binary companion is negligible. The planet reaches an orbit of
$e_p = 0.927$ and $a_p = 0.47$ AU $\sim 1.2$ Gyrs after its birth.
 
A few notes are in order.

We adopt a value of $Q_p = 3 \times 10^5$ for the planet, and $Q_\star
= 10^6$ for the host star.
We discuss the reasons for these choices in \S
\ref{subsec:Qvalues}. Using them, the planet's spin reaches
synchronicity with the orbital frequency near periastron almost
instantly, while the star's spin is hardly affected by tides (see
lower-right plot of Fig. \ref{fig:eggleton_short}). Tidal
circularization is dominated by dissipation in the planet over that in
the star by a factor of $\sim 10$. Most of the dissipation, which
conserves orbital angular momentum, occurs when the planet has been
driven to maximum eccentricity by the companion star.

The results do not depend on the value for the longitude of the
ascending node. This is because in the quadrapole approximation for
secular interactions, the companion orbit is effectively taken to be
circular so there is no preferred axis in the binary plane. This also
explains why the vertical angular momentum of the planet orbit (Kozai
integral) is conserved -- there is a rotational invariance with
respect to the $z$ axis.

The results also do not depend on the value for the argument of
pericentre, $\omega_p$. Kozai oscillations can cause either
circulation or libration in the value of $\omega_p$. Fig. 3 of Holman
\etal (\cite{holman}) shows that the circulating solutions can reach
higher $e_p$ and therefore should lead to shorter evolutionary
timescale. However, for the initial conditions we adopted, the
circulating and librating solutions essentially coincide and give rise
to similar values for the maximum eccentricity.



%
To migrate the planet from a distance larger than $5$AU, one requires
a higher minimum inclination so as to reach the same pericentre
distance at maximum eccentricity. In comparison, this inclination
hardly varies when the initial eccentricity of the planet is varied
from $0.1$ to $0.6$, this can be seen from
Fig. \ref{fig:eggleton_short}.  Lastly, if the companion is further
away than $1000$ AU, each Kozai cycle lasts longer, and it takes a
longer time for the planet to migrate to its current location. For
instance, placing the companion at $a_c = 1500$ AU, we find the planet
should now be $\sim 5 $Gyrs old.
%


\section{Discussion}
\label{sec:discussion}

\subsection{$Q$ Value}
\label{subsec:Qvalues}

A time-dependent tide raised on a celestial body, due to either its
asynchronous rotation (asynchronous tide) or its eccentric movement
(eccentricity tide), can be dissipated, leading to synchronous
spin or orbital circularization. The rate of dissipation is
conventionally described by a dimensionless quality factor $Q$, 
the ratio between the energy in the tidal bulge and the energy
dissipated per orbital period. For equilibrium tides, $Q$ is related to the
lag-angle between the tide-inducing object and the tidal bulge
$\delta$ by $\delta \approx 1/2Q$.

It is possible to adapt the $Q$ description to dynamical tides, in
which case the tidal energy refers to the energy in the (non-existent)
equilibrium tidal bulge, and $Q$ is obtained from the dissipation
averaged over a range of tidal frequencies. Back-reaction to the
orbits by the gravitational moments of the tidal waves is ignored, a
procedure strictly justifiable only when the damping time of the waves
is shorter than the orbital period. However, secular effects of the
tidal dissipation should be independent of this back-reaction.

Jupiter's $Q$ value has been estimated to be $10^5 \la Q \la 2\times
10^6$, with the actual value believed to be closer to the lower limit
(Goldreich \& Soter \cite{goldreich}, Peale \& Greenberg
\cite{peale}), based on the resonant configuration of the Galilean
satellites. The physical origin of this $Q$ value has remained elusive
for a few decades. As is explicitly shown in Wu (\cite{wu}),
exo-planets share similar $Q$ values ($Q_p \sim 3\times 10^5$) as
Jupiter.\footnote{This statement is independent of the $Q_\star$ value
of the host stars. The stars are typically slowly rotating so their
tidal dissipation have the effects of both circularizing and eroding
the planet orbit. This would fail to produce the observed circular
orbits, if it is more important than dissipation in the planets.}
This is striking as the exo-solar planets have different thermal
environments, formation histories, and possibly interior compositions,
than those of Jupiter. The similar $Q$ values provide an important
clue as to the underlying dissipation mechanism.  Moreover, while the
$Q$ factor for Jupiter pertains to the asynchronous tide, the one for
exo-planets concerns the eccentricity tide.  One may speculate that
the quality factors for the different Jovian-mass objects agree
because all the objects rotate fast with periods comparable to or
shorter than the tidal period.

Tidal dissipation in the host star of HD $80606$b also contributes to
circularization.  What is the tidal $Q$ factor appropriate for
solar-type stars? Field solar-type binaries are observed to be
circularized out to $\sim 12$ day orbits (Duquennoy
\& Mayor \cite{duqu}). This would imply $Q \sim 10^5$ adopting an age of $5$
Gyrs.  These binary stars have presumably been synchronized, so this
$Q$ value is affiliated with dissipation of the eccentricity tide in
rapidly spinning objects ($Q_{\rm fast}$). For the evolution of planet
orbits, the $Q$ value of relevance is that for the asynchronous tide
in slowly rotating objects to which most host stars belong ($Q_{\rm
slow}$). It is not clear that $Q_{\rm fast}$ is directly related to
$Q_{\rm slow}$.  What observational constraints could we put on
$Q_{\rm slow}$?

Short period planets can tidally spin up the star, at the cost of
planetary orbital decay. Assuming all host stars are Sun-like and all
exo-planets are Jupiter-like, we obtain a lower limit of $Q_{\rm slow}
> 10^5$ if all planets are to survive in their current orbit for
another $5$ Gyrs.

Drake \etal (\cite{drake}) considered the rotational velocities for
three planet-bearing stars ($\tau$-Boo, $\upsilon$-And, and 51
Peg). Among them, $\tau$-Boo has the closest and most massive
planet and appears to be synchronized,
while the other two have not. Taken at face value, this would imply
$Q_{\rm slow} \approx 10^4$. This contradicts our earlier statement
that $Q_{\rm slow} > 10^5$. These two results can be reconciled either
by the imminent death of the shortest period planets,\footnote{If this
possibility turns out true, tidal dissipation in stars will be held
accountable for the observed inner cutoff in the orbits of
exo-planets.} or by the possibility that $\tau$-Boo has had only its
surface convective layer synchronized with the orbit (see Goldreich \&
Nicholson \cite{goldreich2} for a similar argument on massive stars).
The latter possibility also makes it difficult to infer $Q_{\rm slow}$
using the spin data of solar-type binaries. It is interesting to
notice that among the three stars discussed by Drake \etal
(\cite{drake}), $\tau$-Boo likely has the thinnest convective
envelope.

Unable to infer hard constraints on $Q_{\rm slow}$, we decide to adopt
$Q_\star = Q_{\rm slow} = 10^6$ in our study. With this choice,
dissipation in the planet is roughly $10$ times more important than
dissipation in the star even for a planet as massive as $M_p = 7.8
M_J$. This choice has the advantage that unless the actual $Q_\star$
is much lower, the results presented here are not qualitatively
affected.

\subsection{Constraints on possible Second Planets}
\label{subsec:secondplanet}

If there existed a second planet in the system, secular
interaction\footnote{We ignore interactions of the mean-motion type as
they are important only for a small phase space.}  between this planet
and HD $80606$b may destroy the Kozai oscillations. Kozai oscillations
require that the precession caused by a second planet acting on HD
$80606$b's orbit be no larger than that produced by the companion star
(Innanen \etal, \cite{innanen}).  We briefly investigate what
constraints this puts on the mass and orbit of the second planet.

We assume the two planets have coplanar orbits and the second planet
(denoted with subscript $s$) lies inward of the orbit of HD
$80606$b. The derivation is similar if it lies outward.
Secular interaction between the two planets alone gives rise to
variations in their eccentricities ($e$) and periapse angles
($\omega$). Adopting a complex variable $I = e\, \exp(i \omega t)$,
and assuming that $e \ll 1$, the first-order secular evolution
equations read (Murray \& Dermott \cite{MD2000}, hereafter MD2000),
\begin{equation}
{{dI_s}\over{dt}} = i A_{ss} I_s + i A_{sp} I_p, \hskip0.2cm
{{dI_p}\over{dt}} = i A_{ps} I_s + i A_{pp} I_p,
\label{eq:linearI}
\end{equation}
where the real coefficients $A_{ij}$ are
\begin{eqnarray}
A_{ss} & = & {{\alpha^2}\over{4}} {{M_p}\over{M_\star}} n_s B_1 ,
\hskip0.2cm 
A_{sp}  =  - {{\alpha^2}\over{4}} {{M_p}\over{M_\star}} n_s B_2,\nonumber \\
A_{ps} & = & - {{\alpha }\over{4}} {{M_s}\over{M_\star}} n_p B_2, \hskip0.84cm 
A_{pp}  =   {{\alpha }\over{4}} {{M_s}\over{M_\star}} n_p B_1.
\label{eq:defineAjk}
\end{eqnarray}
Here, $n_s = \sqrt{G(M_\star + M_s)/a_s^3}$, $\alpha = a_s/a_p$, and
$B_1 = b_{3/2}^{(1)}(\alpha)$, $B_2 = b_{3/2}^{(2)}(\alpha)$ with
$b_i^{(j)}$ being the usual Laplace coefficients (MD2000).
The general solution to eq. \refnew{eq:linearI} is composed of two
linear eigenvectors with their respective precession rates
\begin{equation}
g_{\pm} = {{(A_{ss}+A_{pp}) \pm \sqrt{(A_{ss}-A_{pp})^2 + 4
A_{sp} A_{ps}}}\over 2}.
\label{eq:solutiong}
\end{equation}
We are able to approximate the precession rate of HD $80606$b under
secular interaction as,
\begin{eqnarray}
{{d\omega_p}\over{dt}}\Bigr|_{\rm sec} & = & 
{\rm Min}(g_+, g_-)\hskip0.5cm {\rm{if\,\, M_p > M_s}}, \nonumber \\
& & {\rm Max}(g_+, g_-)\hskip0.5cm {\rm{if\,\, M_p < M_s}}.
\label{eq:secularrate}
\end{eqnarray}

\begin{figure*}[t]
\centerline{\psfig{figure=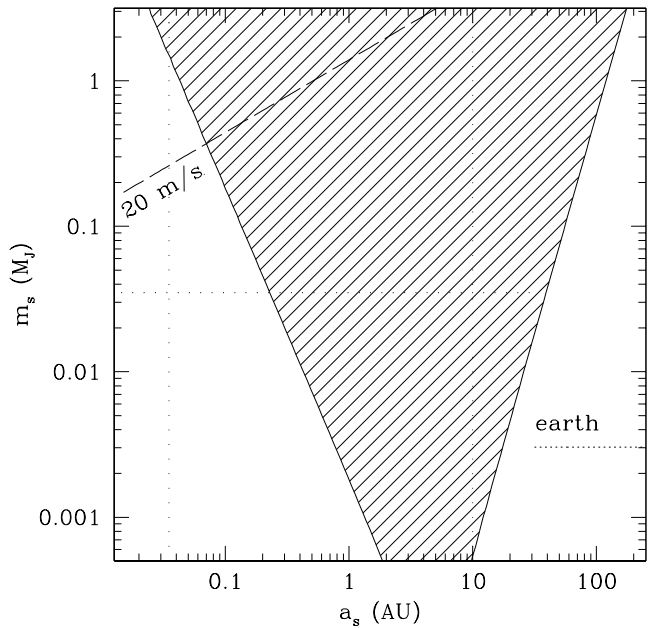,width=0.50\hsize}}
\caption{The constraints on the mass (vertical axis, in unit
of $M_J$) and orbital semi-major axis (horizontal axis, in AU) of the
possible second planet placed by the Kozai migration of HD
$80606$b. The shaded region must be cleared before HD $80606$b
(assumed to be at $5$ AU) can undergo Kozai oscillations. The latter
force HD $80606$b into a variable but highly eccentric orbit with the
minimum periastron and maximum apastron delineated by the two dotted
vertical lines (see Fig. \ref{fig:eggleton_short}).  No planets should
have survived within this region. Planets lying above the slanted
dashed line would incur a radial velocity larger than $20 \m/s$ on the
host star (taking $\sin i=0.5$) and should have been detected were
they present.
}
\label{fig:gaussian_2ndplanet} 
\end{figure*}

At birth, HD $80606$b could not have undergone Kozai oscillation if
the second planet is massive enough.  Requiring the above secular
precession rate to be smaller than the Kozai precession rate (see
eq. [\ref{eq:precession}]) when $a_p = 5$ AU, we find that the mass of
the second planet is constrained to a value that depends on its
distance to HD $80606$b. This is depicted in
Fig. \ref{fig:gaussian_2ndplanet}.

These results also strongly constrain the current state of the
planetary system. No other Jupiter-mass planet could now live between
about $0.05$ and $100$ AU.  Even earth-mass cores are excluded between
$1$ and $20$ AU. Did such objects never form?  Or were they cleared
away? We do not explore these issues here.

In any case, Kozai migration is incompatible with the notion that the
claimed residual velocity in HD$80606$ (NLM) is due to a second planet
in the system.


\subsection{Can the Tide Raised by the Planet Account for the Residual Velocities?}
\label{subsec:tidalheight}

NLM reported residual velocities of order $\sim 40 \m/\s$ in the
spectra of HD $80606$. These are higher than the expected noise and
there is no clear trend in time. Could they be caused by anything
other than an unknown planet? We consider one option: the tidal
velocities induced by HD $80606$b on the surface of the star. The
stellar rotation is ignored in the following analysis.

First, we assume that the star adjusts its hydrostatic equilibrium
instantaneously according to the tidal potential of the planet at a
radial separation $D = D(t)$ (the equilibrium tide picture), and
estimate the corresponding tidal velocities. The maximum radial
displacement at the stellar surface is roughly
\begin{equation}
\xi_{\rm eq} \approx {{M_p}\over {M_\star}} \left({{R_\star}\over D}\right)^3 R_\star,
\end{equation}
with the horizontal displacement being of a similar order.  Velocities
associated with these displacements are simply estimated as $v_{\rm
equi} = {{d\xi_{\rm eq}}\over{dt}}$, yielding a maximum line-of-sight
velocity $\sim 60 {\rm cm/s}$ occurring within $10$ hours of the
closest-encounter. This maximum requires conjunctions to coincide
roughly with the line of periapse, as is the case in HD $80606$.

We turn next to consider a more realistic description of the tidal
amplitudes.  The stellar response toward the time-varying potential
is decomposed into a series of eigenmode oscillations, with the mode
amplitudes maximized when the tidal forcing frequencies are in
resonance with the mode frequencies (the dynamical tide picture). If
the planetary orbit is nearly circular, the dominant forcing frequency
is $\sim 2 n_p$.
At an orbital period of $111$ days, this forcing resonantly excites
gravity-modes of radial order $\sim 1000$. The spatial overlap between
the tidal potential and such a high-order mode is too weak to be
interesting. However, HD $80606b$ has a highly eccentric orbit. Tidal
excitation occurs primarily near periastron and the dominant forcing
frequencies is about twice the orbital frequency near periastron, and
is related to $n_p$ as $\sim 2 n_p/ (1-e_p)^{3/2}
\sim 100 n_p$.  Gravity-modes of radial orders $\sim 30 - 80$ become
relevant and their resonant excitation give rise to much larger tidal
velocities.  A brief treatment of the dynamical tides is given in the
appendix (\S \ref{sec:dyntide}). Here we summarize our results.

\begin{figure*}[t]
\centerline{\psfig{figure=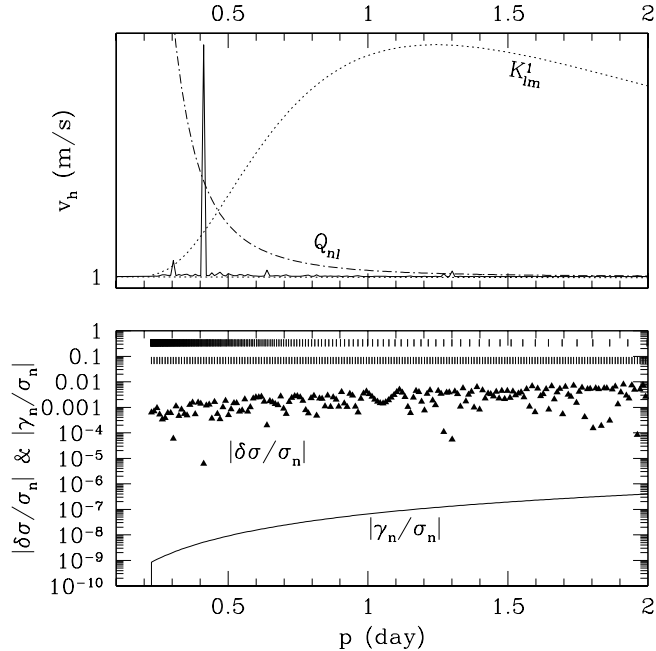,width=0.50\hsize}}
\caption{Dynamic response of a Sun-like star  under tidal forcing from a 
planet of mass $3.9/\sin i M_J$, an orbit of period $111.82$ days and
eccentricity $0.927$. We only consider the dominant $\ell = |m| = 2$
component in the tidal forcing.  The upper panel plots mode amplitude
(in terms of surface horizontal velocity) as a function of mode period
in days. Usually one or two modes dominate the response, and in this
case it is a $n= 41$ gravity-mode. The dotted and dashed lines are the
frequency spectrum of tidal forcing over one period ($K^1_{\ell m}$,
see appendix), and the overlap integral between the tidal potential
and the eigenmode ($Q_{n\ell}$), both in arbitrary units. The forcing
spectrum peaks around $2 (1-e_p)^{-3/2} n_p$ and the overlap integral
decreases sharply when mode period (and mode radial order) increases,
roughly as $Q_{n \ell} \propto n^{-5}$. Besides from these two
functions, the tidal response also depends on the goodness of the
resonance and the mode damping rate, these are plotted in the lower
panel.  The frequency detuning $\delta \sigma = |\sigma_n -
\sigma_j|$. Here $\sigma_n$ is the frequency of a radial order $n$ mode, 
and $\sigma_j$ is the closest integer multiple of the orbital mean
motion ($\sigma_j = j n_p$ and $j$ is an even number).  The damping
rate $\gamma_n$ is estimated from radiative diffusion.  The two rows
of short vertical bars illustrate the positions of the $\sigma_j$
(upper row) and $\sigma_n$ (lower).}
\label{fig:dynamic_tide} 
\end{figure*}

We adopt the standard solar model (Christensen-Dalsgaard
\cite{jorgen}) to calculate the eigenmode frequencies and damping rates.
The response of the star depends sensitively on how close an orbital
multiple lies around one such eigenmodes. The maximum response occurs
when the frequency off-resonance ($\delta \sigma$) is of order or
smaller than the mode damping rate ($\gamma$), giving rise to
velocities of order $10$ to $100 \m/\s$. However, this is a rare
event. In the frequency range of interest ($150$ to $550 n_p$), there
are roughly $N_{\rm peak} \sim 200$ discreet forcing peaks, and
$N_{\rm mode} \sim 50$ eigenmodes. These modes typically have
line-widths calculated from radiative diffusion to be
$\gamma_n/\sigma_n \sim 10^{-7}$. A typical best resonance
$\delta\sigma = {\rm Min}(|\sigma_n - \sigma_j|)$ is of order $\sim
n_p/N_{\rm mode} \sim \sigma_n/10^4 \sim 10^3 \gamma_n$. So the chance
of one mode lying close to an orbital multiple with $\sigma_n
\leq \gamma_n$ is roughly $10^{-3}$. Typical tidal velocities therefore
range from $1\cm/s$ to $10 \cm/s$, well below the level of the
observed residues. One such example is shown in
Fig. \ref{fig:dynamic_tide}. The highest amplitude mode has a
fractional Lagrangian pressure perturbation $\delta p/p \sim 4\times
10^{-7}$ near the surface. Translating this into light variation, this
mode will be as hard to detect as the solar oscillation.

\begin{figure*}[t]
\centerline{\psfig{figure=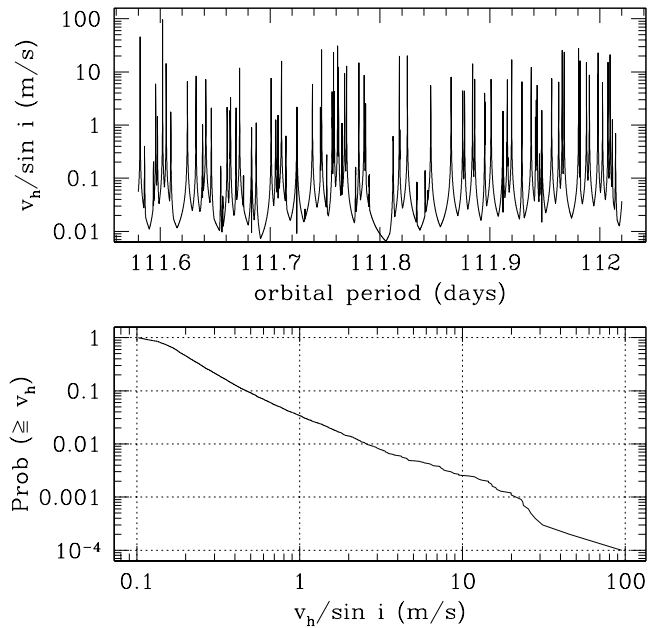,width=0.50\hsize}}
\caption{The probability distribution of the tidal response. 
The upper panel plots the horizontal velocity on the stellar surface
as a function of orbital period, with the period running through the
observational uncertainty range. Sharp spikes occur when an eigenmode
is resonantly forced ($\delta \sigma < \gamma_n$).  Other parameters
are the same as in Fig. \ref{fig:dynamic_tide}. From these data, one
finds a probability distribution for the tidal velocity to surpass a
certain value. This is depicted in the lower panel.  Typical amplitude
(probability of $0.5$) is $\sim 20 \cm/s$, with the claimed level of
residue, $\sim 40\m/\s$, occurring at a probability of $\sim 10^{-3}$.}
\label{fig:probable_tide} 
\end{figure*}

To showcase the sensitivity of the tidal responses, we plot in
Fig. \ref{fig:probable_tide} the marked variations in the tidal
responses when the orbital period is tuned through the observed
uncertainty range: $111.81 \pm 0.23$ days (NLM). Slight differences in
the stellar structure affect the eigenmode frequencies and would also
give rise to similar variations in the tidal responses. So the
response results are best appreciated as a probability distribution of
amplitudes, also shown in Fig. \ref{fig:probable_tide}. We find that
there is only a $\sim 10^{-3}$ chance that the tidal velocity reaches
the claimed level of residues, $\sim 40\m/\s$.

We calculate the damping rates taking only radiative diffusion into
account. Turbulent viscosity may enhance this damping. But this will
not appreciably change our conclusion: raising the damping rates by a
factor of $10$, we obtain similar results as in
Fig. \ref{fig:probable_tide} except that high tidal amplitudes are
less likely to occur.

On the other hand, when we substitute the solar model for a $1.25
M_\odot$ ZAMS model (courtesy of Christensen-Dalsgaard), we found that
the typical tidal amplitudes rise by about a factor of
$10$. Equivalently, high velocity amplitudes appear roughly $10$ times
more frequent. Compared to a solar model, the $1.25 M_\odot$ model has
a thinner convection zone. Its gravity-modes have a relatively larger
surface amplitude.

In summary, thanks to the high eccentricity of HD $80606$b, the tidal
velocities reach tantalizingly close to the current detection
thresholds, but likely fail to explain any residual velocity of order
$10 \m/\s$. However, if the residual velocity is indeed associated
with the dynamical tide, we expect it to be a coherent oscillation
throughout the orbit and have a period in the range of $0.2$ to $0.6$
days. The inclination of the orbit ($\sin i$) does not affect our
conclusions.

\subsection{Other Planetary Systems}
\label{subsec:othersystem}

\begin{figure*}[t]
\centerline{\psfig{figure=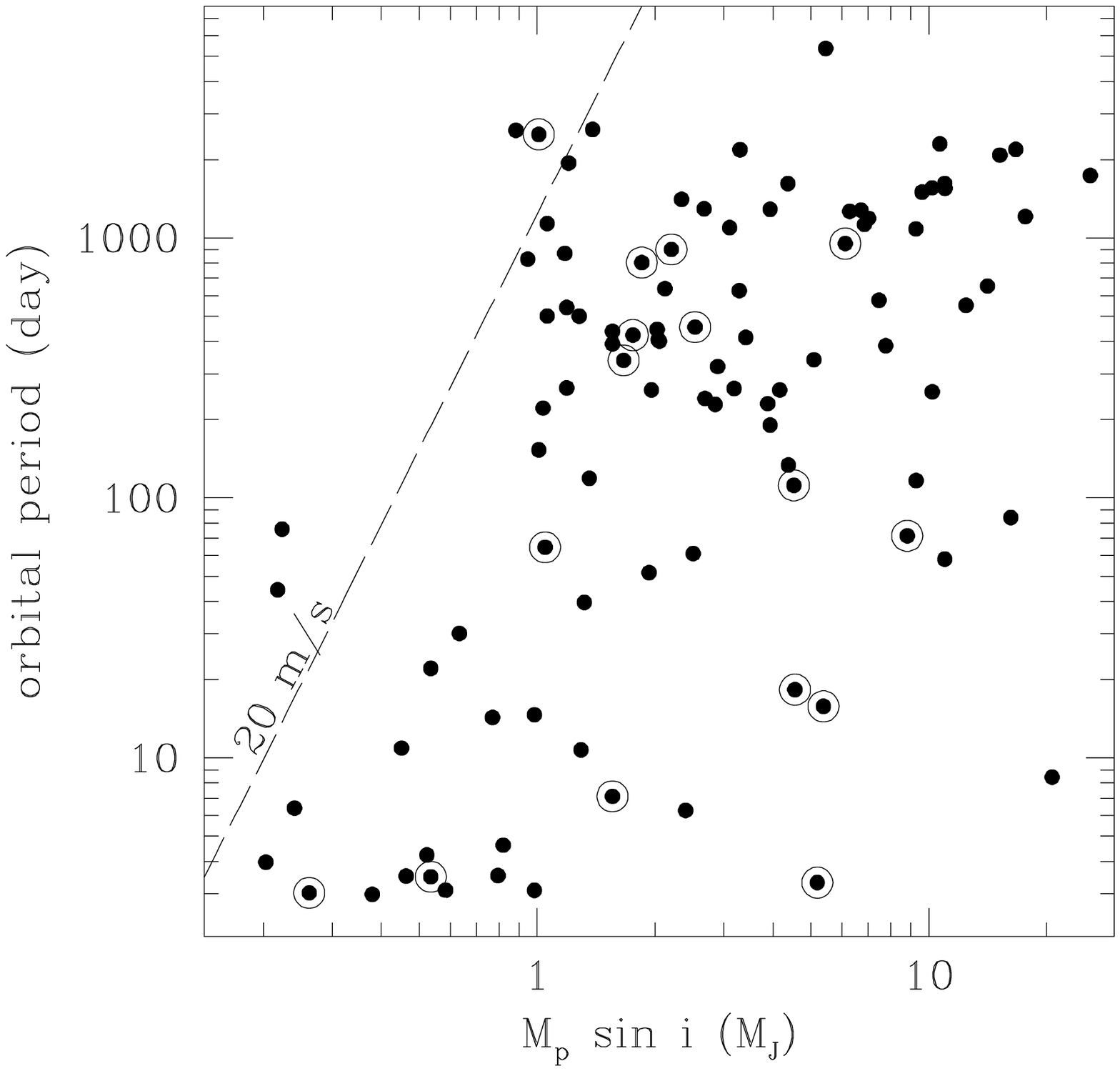,width=0.50\hsize}}
\caption{Period versus projected mass for all observed planet 
candidates (http://cfa-www.harvard.edu/planets/, Sep. 2002). Circles
are drawn around planets in visual binaries (Worley \& Douglass
\cite{worley}) with a projected separation $< 20\arcsec$. Besides from
the 9 binaries mentioned in Zucker \& Mazeh (\cite{zucker}, HD 142, Gl
86, HD 19994, $\epsilon$ Eri, $\tau$ Boo, HD 178911B, 16 Cyg B, HD
195019, HD 217107), there are also HD 80606, HD 213240, HD 121504, HD
46375, HD 27442, BD-103166, and $\gamma$-Cepheid.
%
%
Planets lying below the dashed line incur velocity variation higher
than $20\m/\s$. 
%
There appears to be an over-abundance of close-in, massive planets
that are also in binaries. 
}
\label{fig:mass_period} 
\end{figure*}

Zucker \& Mazeh (\cite{zucker}) showed that planets in binaries are
statistically different from those around isolated stars in their
period-mass distribution. The former population dominates the
lower-right quadrant of Fig. \ref{fig:mass_period}.  This distinction
suggests that planet migration in binaries may be induced differently
from those around single stars, Kozai migration being one such
possibility. Among the circled planets in Fig. \ref{fig:mass_period},
we have only studied $\tau$-Boo and 16 CygB. We found that $\tau$-Boo,
the most massive close-in planet, could have gone through Kozai
migration and later circularized, while there is no compelling
evidence of Kozai migration for 16 CygB, though it could currently be
undergoing Kozai oscillations (Holman \etal (\cite{holman}). A binary
companion may assist planet migration through other means, e.g.,
exciting tidal waves in the proto-planetary disk and affecting angular
momentum transport. These deserve further study.

Could some of the planets around currently single stars have undergone
Kozai migration followed by the subsequent disruption of the binary? 
We consider this unlikely. The rate of Kozai migration is limited by
the tidal circularization process. For reasonable values of closest
periapse distance, the latter process takes a few $10^8$ years or
longer. By this age a typical stellar cluster has dispersed and the
%
rate of stellar encounter has become negligible. However, it is
possible that a planet's eccentricity was excited by Kozai interaction
with the binary companion before the latter was removed from the
system. A simple estimate, however, finds that this may account for
the observed eccentricity in at most $\sim 10\%$ of the known systems.

\section{Summary}
\label{sec:summary}

It is striking that a companion star as remote as $1000$ AU can force
a planet to migrate inward. Kozai migration is a long timescale
process and demands a 'clean' planetary system 
in which no other planetary object exists.
Kozai migration can be responsible for the
present orbit of HD $80606$b only if the initial relative inclination
is sufficiently high ($I \geq 85 \deg$). Assuming
a population of similar triple systems with randomly distributed $I$,
$\sim 7\%$ of them could have gone through Kozai migration.  So on the
strength of one object (HD $80606$b), one may speculate about the high
density of triple stellar-planetary systems yet undetected. 


Our prediction regarding the lack of a second planet in the system
(Fig. \ref{fig:gaussian_2ndplanet}) should be tested. In this regard,
we note the claim of the abnormally large residue velocities on the
star.  We investigated the possible association between these
velocities and surface tidal waves excited by the highly eccentric
planet. It seems unlikely that the latter can account for the former
if the star has a structure like that of our Sun. And if the system is
indeed devoid of second planet, it raises issues as to how it is
cleaned.

Zucker \& Mazeh (\cite{zucker}) pointed out that planets in binary
systems and planets around single stars do not share the same
period-mass distribution. The implications might be that binaries can
cause planet migration -- Kozai migration being one of the more
definite possibilities.

HD $80606$b's highly elongated orbit and strong tidal interaction
combine to make it a possible target for direct detection. Its current
tidal luminosity (which is likely to be converted into heat and
radiated all through the orbit) is $\sim 10^{28} \erg/\s$, or $\sim
14$ mag dimmer than the parent star, at a maximum angular separation
of $0.01\arcsec$ (assuming distance of $80$ pc). A comparable
contribution may arise if the planet evenly emits the stellar
insolation it receives during periastron passages.

\begin{acknowledgements}
Part of the research was performed while Y.W. was on a CITA-ITP
postdoctoral exchange programme funded by NSF grant PHY 99-07949.
Y.W. would like to thank Peter Eggleton for his patience, Ludmila
Kiseleva-Eggleton, Omer Blaes and Man-Hoi Lee for useful
conversations. N.M.'s research is supported by NSERC of Canada and by
the Canada Research Chair program.  This research benefited from ADS
and SIMBAD.
\end{acknowledgements}

{}

\begin{appendix}

\section{Equations from EKE}
\label{sec:EKEeqn}
	
We adopt the set of equations (eq. [11]-[17]) in EKE (also see
Eggleton, Kiseleva \& Hut \cite{EKH}) to describe the secular
evolution in the planet orbit, the spin of the host star and the
planet. The companion is assumed to be unaffected. Physical effects
described by these equations include: secular interactions, GR, tidal
and rotational precessions, and tidal dissipation. These equations are
explicitly listed here,
\begin{eqnarray}
{1\over{e_p}}{{de_p}\over{dt}} & = & - V_\star - V_p - 5 (1-e_p^2)
S_{eq}, \label{eq:ep} \\ 
{1\over{a_p}}{{da_p}\over{dt}} & = & - 2 W_\star -
2 W_p - {{2 e_p^2}\over{1-e_p^2}} (V_\star + V_p), \label{eq:ap} \\
{{dI}\over{dt}} & = & - \sin \omega_p \left[Y_\star + Y_p + (1-e_p^2)
S_{qh}\right] + \cos\omega_p \left[ X_\star + X_p + (4 e_p^2 +1)
S_{eh}\right], \label{eq:i} \\
{{d\omega_p}\over{dt}} & = & \left[Z_\star+ Z_p + Z_{\rm GR} + (1-e_p^2)
(4 S_{ee} - S_{qq})\right] - \cos \omega_p {{\cos I}\over{\sin I}}
\left[ Y_\star + Y_p + (1-e_p^2) S_{qh}\right] \nonumber \\
& &  - 
\sin \omega_p {{\cos I}\over{\sin I}} \left[X_\star + X_p + 
(4 e_p^2 +1) S_{eh}\right], \label{eq:smallomega} \\
{{d\Omega}\over{dt}} & = & {{\cos\omega_p}\over{\sin I}}\left[ Y_\star +
Y_p + (1-e_p^2) S_{qh}\right] + {{\sin \omega_p}\over{\sin I}} \left[
X_\star + X_p + (4 e_p^2 + 1) S_{eh}\right], \label{eq:bigomega} \\
{{d \Omega_{* \gamma}}\over{dt}} & = & {{\mu h_p}\over{I_\star}} ( -
Y_\star e_\gamma + X_\star q_\gamma + W_\star h_\gamma), \label{eq:omegasi} \\
{{d \Omega_{p \gamma}}\over{dt}} & = & {{\mu h_p}\over{I_p}} ( - Y_p e_\gamma +
X_p q_\gamma + W_p h_\gamma). \label{eq:omegapi}
\end{eqnarray}
In eq. \refnew{eq:omegasi} - \refnew{eq:omegapi}, the subscript
$\gamma$ runs through the three spatial indexes, $E$, $Q$, and $H$,
with $\hat E$, $\hat Q$, $\hat H$ being the three axis defining the
orbital plane of the binary companion, and $\hat e$, $\hat q$ and
$\hat h$ defining that of the planet orbit, and $e_E = \hat e \cdot
\hat E$, etc. The inclination $I$ is taken to be that between the two
planes, with the binary plane held constant during the evolution. The
other two angles, $\omega_p$ and $\Omega$, are the argument of the
planet pericentre, and the longitude of the planet ascending node,
respectively.  The spin vector $\Omega_\star$ ($\Omega_p$) belongs to
the host star (planet), and $I_\star = r_{g\star} M_\star R_\star^2$
is the stellar moment of inertia, with $r_{g\star}$ typically called
the gyro-radius. $I_p$ is for the planet.  The planet orbit has an
angular momentum of magnitude $h_p = [G (M_\star + M_p) a_p
(1-e_p^2)]^{1/2}$, and the reduced mass of the planet-star system is
$\mu = M_\star M_p/(M_\star +M_p)$.  All other symbols are defined in
EKE. We do not repeat these definitions here.

We introduce two commonly used dimensionless numbers, $k_{2}$ (the
tidal Love number) and $Q$ (the tidal quality factor), and relate them
to the notation in EKE,
\begin{eqnarray}
A_\star = k_{2\star} R_\star^5, &  & A_p  =  k_{2p} R_p^5, \label{eq:defA}\\
{1\over{t_{F\star}}}  = {{3 k_{2\star}}\over{Q_\star}}
\left({{R_\star}\over{a_p}}\right)^5 {{M_p}\over{M_\star}}\, n_p, & & 
{1\over{t_{Fp}}}  =  {{3 k_{2p}}\over{Q_p}}
\left({{R_p}\over{a_p}}\right)^5 {{M_\star}\over{M_p}}\, n_p . \label{eq:deftF}
\end{eqnarray}

\section{Dynamical Tide}
\label{sec:dyntide}

We provide a simple description of the dynamical tide treatment. The
final results are presented in \S \ref{subsec:tidalheight}.

The period of the stellar spin is likely long compared to both the
sound crossing time and the dominant tidal forcing period. So we
assume a non-rotating star.  Following Press \& Teukolsky
(\cite{press}) and Lai (\cite{donglai}) and notations therein, we
decompose the tidal potential at a point ${\bf r} = [r,\theta,\phi]$
inside the star, due to a planet at location ${\bf D}(t) = [D(t),
\pi/2, f(t)]$, as
\begin{equation}
U ({\bf r}, t) = - G M_p \sum_{\ell, m} W_{\ell m}
{{r^{\ell}}\over{D^{\ell+1}}} Y_{\ell m} (\theta,\phi) \exp^{-i m
f(t)}.
\label{eq:tidepotential}
\end{equation}
The stellar fluid responds with a displacement ${\bf \xi}({\bf r},t)$
which can be projected onto various free eigen-oscillations,
\begin{equation}
{\bf \xi}({\bf r}, t) = \sum_\alpha a_\alpha(t) {\bf \xi}_\alpha ({\bf
r}).
\label{eq:atalpha}
\end{equation}
The eigenfunctions are normalized as $\int d^3 x \rho {\bf
\xi}_\alpha \cdot {\bf \xi}_{\alpha}^* = 1$.
We denote the mode eigenfrequency as $\sigma_\alpha$ and damping rate
as $\gamma_\alpha$, and introduce ${\tilde a}_\alpha (\sigma)$ which
is related to $a_\alpha (t)$ by
\begin{equation}	
\hskip1.0cm a_\alpha (t) = \int_0^{\infty} d\sigma \exp^{-i\sigma
t} {\tilde a}_\alpha (\sigma).
\label{eq:ataa}
\end{equation}
Substituting equation \refnew{eq:atalpha} into the fluid equation of
motion, we obtain
\begin{equation}
{\tilde a}_\alpha(\sigma) = {1\over{\sigma_\alpha^2 - \sigma^2+ 2 i
\gamma_\alpha \sigma}} \sum_\ell {{G M_p}\over{D_p^{\ell+1}}}
Q_{\alpha\ell} K_{\ell m} (\sigma),
\label{eq:aalpha}
\end{equation}
where the periapse distance $D_p = a_p (1-e_p)$, $K_{\ell m}(\sigma)$
is the frequency spectrum for the $Y_{\ell m}$ component of the tidal
potential, and $Q_{\alpha \ell}$ is the spatial overlap integral
between the tidal forcing and the $\alpha$ eigenfunction,
\begin{eqnarray}
K_{\ell m}(\sigma) & = & {{W_{\ell m}}\over{2
\pi}} \int^{\infty}_{0} dt \left[{{D_p}\over{D(t)}}\right]^{\ell +1}
\exp[- im f(t) + i\sigma t],\label{eq:defineK} \\
Q_{\alpha \ell} & = & \int d^3 x\,\, \rho {\bf \xi}_\alpha^* \cdot
\nabla(r^\ell Y_{\ell m}) = \int d^3 x\,\, \rho_\alpha^{\prime *}\, r^\ell
Y_{\ell m}.
\label{eq:defineQ}
\end{eqnarray}
Here, $\rho_\alpha^\prime$ is the Eulerian density perturbation for
mode $\alpha$. The continuity equation gives $\nabla\cdot(\rho
\xi_\alpha) = - \rho_\alpha^\prime$.

We first focus on $K_{\ell m}$. For a strictly periodic orbit,
$K_{\ell m}(\sigma)$ equals a sum of delta functions with non-zero
values only at the integer multiples ($j$) of the orbital mean
motion. 
\begin{equation}
K_{\ell m}(\sigma) = \sum_{j=0}^\infty C_j(\sigma)\, \delta(\sigma - j
n_p)
\label{eq:Klminf}.
\end{equation}
To obtain $C_j$, we calculate the tidal spectrum integrated over one
orbital period,
\begin{equation}
K^1_{\ell m}(\sigma) = {{W_{\ell m}}\over{2 \pi}}
\int_0^{2\pi/n_p} dt \left[{{D_p}\over{D(t)}}\right]^{\ell +1}
\exp[-im f(t) + i\sigma t].
\label{eq:defineK0b}
\end{equation} 
Unlike $K_{\ell m}$, this quantity is a smooth function of frequency.
For $\ell = |m| = 2$, the forcing maximum lies around $\sigma \sim
(1-e_p)^{-3/2} |m| n_p$, i.e., $\sigma \sim 2 n_p$ for $e_p \ll 1$ and
$\sigma \sim 90 n_p$ for $e_p = 0.927$. This is shown in Figure
\ref{fig:dynamic_tide}. When integration in $K_{\ell m}$
is carried out over $N$ number of orbital cycles, the resulting
forcing spectrum can be approximately described by the following
log-normal distribution,
\begin{equation}
K^N_{\ell m}(\sigma) = \sum_{j=1}^{\infty} N 
\exp\left[-({\sigma\over{n_p}} - j )^2 N^2\right] \, K^1_{\ell m}(\sigma).
\label{eq:lognormal}
\end{equation}
And when $N \rightarrow \infty$, we retrieve $K_{\ell m}$ with
$C_j(\sigma) = \sqrt{\pi} n_p K_{\ell m}^1 (\sigma)$

Substituting the above expression for $K_{\ell m}(\sigma)$ into
equations \refnew{eq:atalpha}- \refnew{eq:aalpha}, we obtain the mode
amplitude,
\begin{equation}
a_\alpha (t) \approx {{G M_p}\over{D_p^{\ell + 1}}} \sqrt{\pi} n_p \,
Q_{\alpha \ell} K^1_{\ell m} (\sigma_j) \int_{(j-{1\over
2})n_p}^{(j+{1\over 2})n_p} d\sigma\, {{\delta(\sigma - j n_p) e^{- i
\sigma t}}\over{\sigma_\alpha^2 - \sigma^2 + 2 i
\gamma_\alpha \sigma}} = {{G M_p}\over{D_p^{\ell + 1}}} \sqrt{\pi} n_p\, Q_{\alpha
\ell}  K^1_{\ell m} (\sigma_j) {{ e^{- i \sigma_j
t}}\over{\sigma_\alpha^2 - \sigma_j^2 + 2 i \gamma_\alpha \sigma_j}},
\label{eq:expressat}
\end{equation}
where $\sigma_j = j n_p$ is the closest orbital multiples to the mode
frequency $\sigma_\alpha$. 


The value of the tidal overlap, $Q_{\alpha \ell}$, decays sharply with
both $\ell$ and the radial order ($n$) of the mode. For a high order
gravity-mode, most of the net contribution to $Q_{\alpha\ell}$ comes
from the upper evanescent region of the mode -- the bottom of the
surface convection zone where $\rho^\prime/\rho$ is fairly
constant. We find that $Q_{\alpha\ell}$ scales with $n$ as
\begin{equation}
Q_{\alpha\ell} \propto n^{- 5}.
\end{equation}






\end{appendix}

\end{document}